\documentclass{emulateapj}

\usepackage[usenames, dvipsnames]{color}

\shorttitle{Buoyant Magnetic Loops in a Global Dynamo Simulation}
\shortauthors{Nelson et al.}

\begin{document}

\title{Buoyant Magnetic Loops in a Global Dynamo Simulation of a Young Sun}

\author{Nicholas J. Nelson\altaffilmark{1}, Benjamin P. Brown\altaffilmark{2}, Allan Sacha Brun\altaffilmark{3}, Mark S. Miesch\altaffilmark{4}, \& Juri Toomre\altaffilmark{1} }

\altaffiltext{1}{JILA and Dept. Astrophysical \& Planetary Sciences, University of Colorado, Boulder, CO 80309-0440}
\altaffiltext{2}{Dept. Astronomy and Center for Magnetic Self-Organization (CSMO) in Laboratory and Astrophysical Plasmas, University of Wisconsin, Madison, WI 53706-1582}
\altaffiltext{3}{Laboratoire AIM Paris-Saclay, CEA/Irfu Universit\'e Paris-Diderot CNRS/INSU, 91191 Gif-sur-Yvette, France.}
\altaffiltext{4}{High Altitude Observatory, NCAR, Boulder, CO 80307-3000}

\begin{abstract}
The current dynamo paradigm for the Sun and sun-like stars places the generation site for strong toroidal magnetic structures deep in the solar interior. 
Sunspots and star-spots on sun-like stars are believed to arise when sections of these magnetic structures become buoyantly unstable and rise from the deep interior to the photosphere. 
Here we present the first 3-D global magnetohydrodynamic (MHD) simulation in which turbulent convection, stratification, and rotation combine to yield a dynamo that self-consistently generates buoyant magnetic loops. 
We simulate stellar convection and dynamo action in a spherical shell with solar stratification, but rotating three times faster than the current solar rate. 
Strong wreaths of toroidal magnetic field are realized by dynamo action in the convection zone. 
By turning to a dynamic Smagorinsky model for subgrid-scale turbulence, we here attain considerably reduced diffusion in our simulation. 
This permits the regions of strongest magnetic field in these wreaths to rise toward the top of the convection zone via a combination of magnetic buoyancy instabilities and advection by convective giant cells. 
Such a global simulation yielding buoyant loops represents a significant step forward in combining numerical models of dynamo action and flux emergence.
\end{abstract}


\section{Convection, Rotation and Magnetism}

The clearest signature of the global solar dynamo is the emergence of sunspots at the photosphere. 
Creating these coherent magnetic structures likely requires several dynamical processes operating at various locations in the solar interior. 
A single 3-D numerical simulation of solar magnetism that extends from the deep interior through the Sun's upper atmosphere, while resolving all relevant scales, is intractable with current computational resources. 
This leads to three main classes of simulations that address elements of solar-like dynamo processes \cite[see reviews][]{Fan09, Cha10}. 
One approach to study how loops may emerge is to insert a compact magnetic field structure into a spherical domain and track its buoyant rise \cite[e.g.,][]{Cal95, Fan08, Jou09, Web11}. 
Another approach uses local planar models with mechanical forcing to generate large-scale shear that drives dynamo action and creates buoyant magnetic loops \cite[e.g.,][]{Cli03, Vas08, Gue11}.  Planar models have also been used to study 3-D buoyancy instabilities in a magnetized layer that can lead to rising elements \cite[e.g.,][]{Ker07}. 
The third approach uses global convective MHD models. 
These incorporate the rotating spherical-shell geometry needed to self-consistently generate differential rotation and meridional circulation through Reynolds stresses \cite[see review][]{Mie05}.
Such models have captured the formation of magnetic structures and cycles in solar \cite{Ghi10, Rac11} and rapidly-rotating sun-like stellar models \cite[][hereinafter B10 and B11, respectively]{Bro10, Bro11b}, yielding differential rotation, dynamo action, and large-scale magnetic fields, but not buoyant magnetic loops that rise toward the top of the convective layer.

\begin{figure*}[]
\begin{center}
\includegraphics[width=\textwidth]{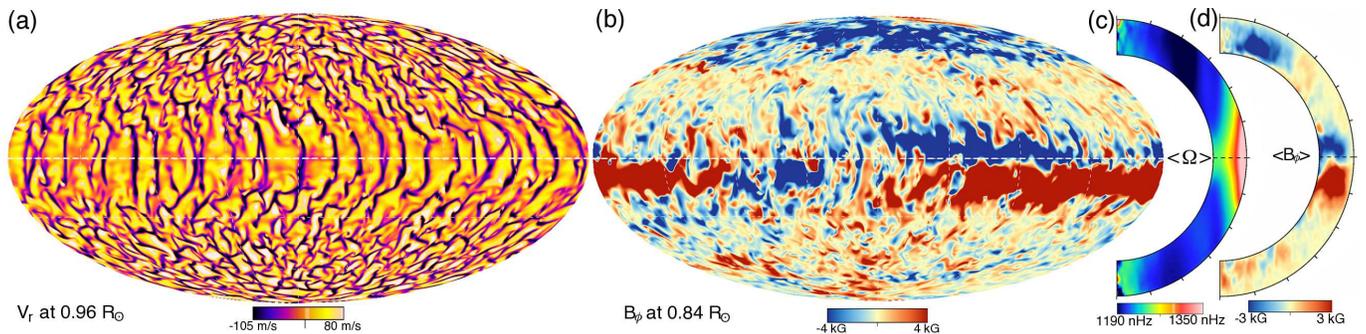}
\caption{ \footnotesize Snapshots of flows and fields in case S3 when a buoyant loop begins to rise at time $t_b$. 
(a)~Radial velocity $v_r$ in global Mollweide projection (equator is dashed) near the top of the computational domain, showing fast, narrow downflows (dark tones) and broad, slower upflows (light tones). 
(b)~Companion view of toroidal magnetic field $B_\phi$ at mid-convection zone. 
Effects of turbulent convection contribute to the ragged nature of the wreaths. 
Several buoyant magnetic loops (see Figure \ref{rise}) are generated in the negative-polarity wreath segment just above the equator and right of image center. 
(c)~Time and zonal average of rotation profile $\langle \Omega \rangle$, possessing an equatorial region with fast rotation and slower rotation at higher latitudes. 
$(d)$~Longitudinally-averaged toroidal magnetic field $\langle B_\phi \rangle$ revealing a prominent axisymmetric field component. 
\label{meetDS}}
\end{center}
\end{figure*}

Here we report on a global convective dynamo simulation of a sun-like star rotating at three times the mean solar angular velocity (3$\Omega_\odot$), such as our Sun did when it was younger and as do many solar analogues \cite{Pet08}. 
This simulation (i) attains a differential rotation profile created by the interplay of convection, rotation and stratification \cite[e.g.,][]{Bru02, Mie09}, (ii) forms global-scale toroidal magnetic structures that undergo cycles of magnetic activity and reversals of global polarity, and (iii) achieves buoyant magnetic loops from the strongest portions of the toroidal structures which rise from the base of the convection zone. 
This work extends the work of B10 and B11 in which simulations of rapidly-rotating suns with moderate levels of diffusion were able to accomplish (i) and (ii). 
The formation of buoyant loops is facilitated in our current work by adopting a dynamic Smagorinsky subgrid-scale model \cite{Ger91}, which serves to minimize the diffusion of well-resolved structures.

\section{Simulation Parameters and Properties}

We have conducted 3-D MHD simulations of turbulent convection and dynamo action in a spherical shell spanning the bulk of the convection zone from 0.72$R_\odot$ to 0.97$R_\odot$ involving a density contrast of 25, and rotating at 3$\Omega_\odot$ (1240 nHz, once every 9.3 days). 
We use the anelastic spherical harmonic (ASH) code \cite[e.g., ][]{Bru04}. 
The anelastic treatment lets us follow the subsonic flows in the deep convection zone. 
Within this nearly adiabatically stratified region, we expect that magnetic buoyancy instabilities captured by our anelastic treatment differ from fully compressible treatments by no more than a few percent in either growth rate or scale \cite{Ber10}. 
ASH is a large-eddy simulation (LES) code that resolves the largest scales of motion and uses a subgrid-scale (SGS) model to parameterize the effects of unresolved, small-scale turbulence. 
The dynamo simulations of B10 and B11 used a SGS model where the turbulent magnetic diffusivity $\eta_t$ was constant on spherical shells and in time, and varied only slowly with depth as the inverse square-root of the background density. 
B10 examined a simulation (case D3, at 3$\Omega_\odot$) which exhibited persistent toroidal magnetic structures, whereas B11 studied a simulation that achieved cycles of magnetic activity and global polarity reversals (case D5, at 5$\Omega_\odot$). 
These simulations had $\eta_t = 2.64 \times 10^{12}$ and $1.88 \times 10^{12} \mathrm{ \, cm}^2 \mathrm{ \, s}^{-1}$ respectively at mid-convection zone.

\begin{figure}[h]
\begin{center}
\includegraphics[width=0.48\textwidth]{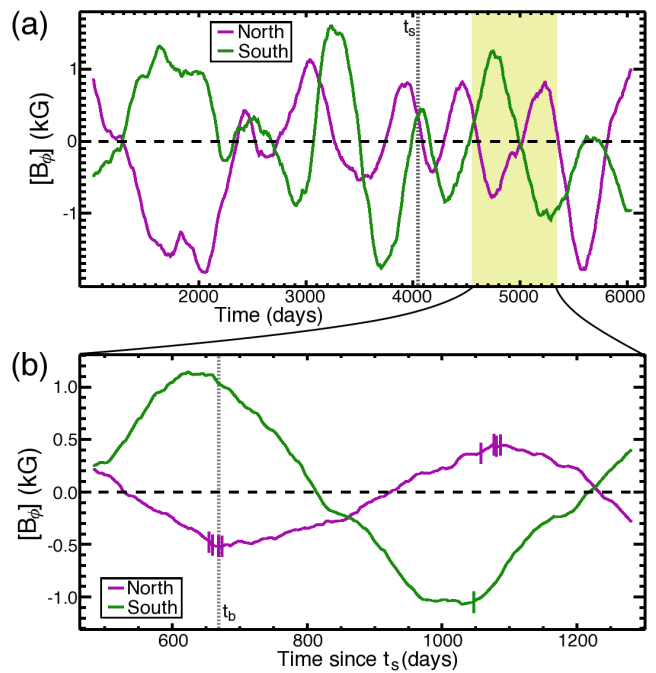}
\caption{ \footnotesize Field reversals with time. 
(a)~Hemispherical volume-averaged toroidal magnetic field $\left[ B_\phi \right]$ of progenitor case D3b over nearly 6000 days, displaying irregular magnetic activity cycles. 
Case S3 branched from case D3b at time $t_s$ (dotted line). 
(b)~$\left[ B_\phi \right]$ for case S3 over about 800 days. 
Case S3 continues the cyclic behavior of D3b, but additionally produces buoyant loops. 
The creation of loops which pass 0.90$R_\odot$ are indicated by tick marks in the lower panel. 
Detailed information on the buoyant loop at time $t_b = t_s + 683$ days (dotted line) is shown in Figures~\ref{meetDS},\ref{rise}.   
\label{history}}
\end{center}
\end{figure}

Here we consider a new ASH simulation, case S3, which achieves much lower levels of diffusion through the use of a dynamic Smagorinksy (DSMAG) SGS model. 
This assumes self-similar behavior in the resolved portion of the inertial range of scales in a turbulent flow in order to extrapolate the effects of unresolved small-scale motions on the resolved scales. 
The resulting viscosity $\nu_S$ is determined by the properties of the grid and the flows, and varies by orders of magnitude in all three spatial dimensions and in time. 
To determine the thermal and magnetic diffusion coefficients we assume constant thermal and magnetic Prandtl numbers. 
In cases D3, D5 and S3 these are set to 0.25 and 0.5 respectively. 
We reserve further discussion of the properties of ASH simulations using the DSMAG SGS model for a forthcoming paper. 
In case S3 the DSMAG SGS model allows a simulation (with 1024 longitudinal, 512 latitudinal, and 193 radial grid points) to achieve a mean magnetic diffusion coefficient at mid-convection zone of $\bar{\eta}_t = 4.8 \times 10^{10} \mathrm{ \, cm}^2 \mathrm{ \, s}^{-1}$. 
This reduction in diffusion by a factor of about 40 from case D3 is critical for the formation and coherent rise of buoyant magnetic loops.

\begin{figure*}[t]
\begin{center}
\includegraphics[width=\textwidth]{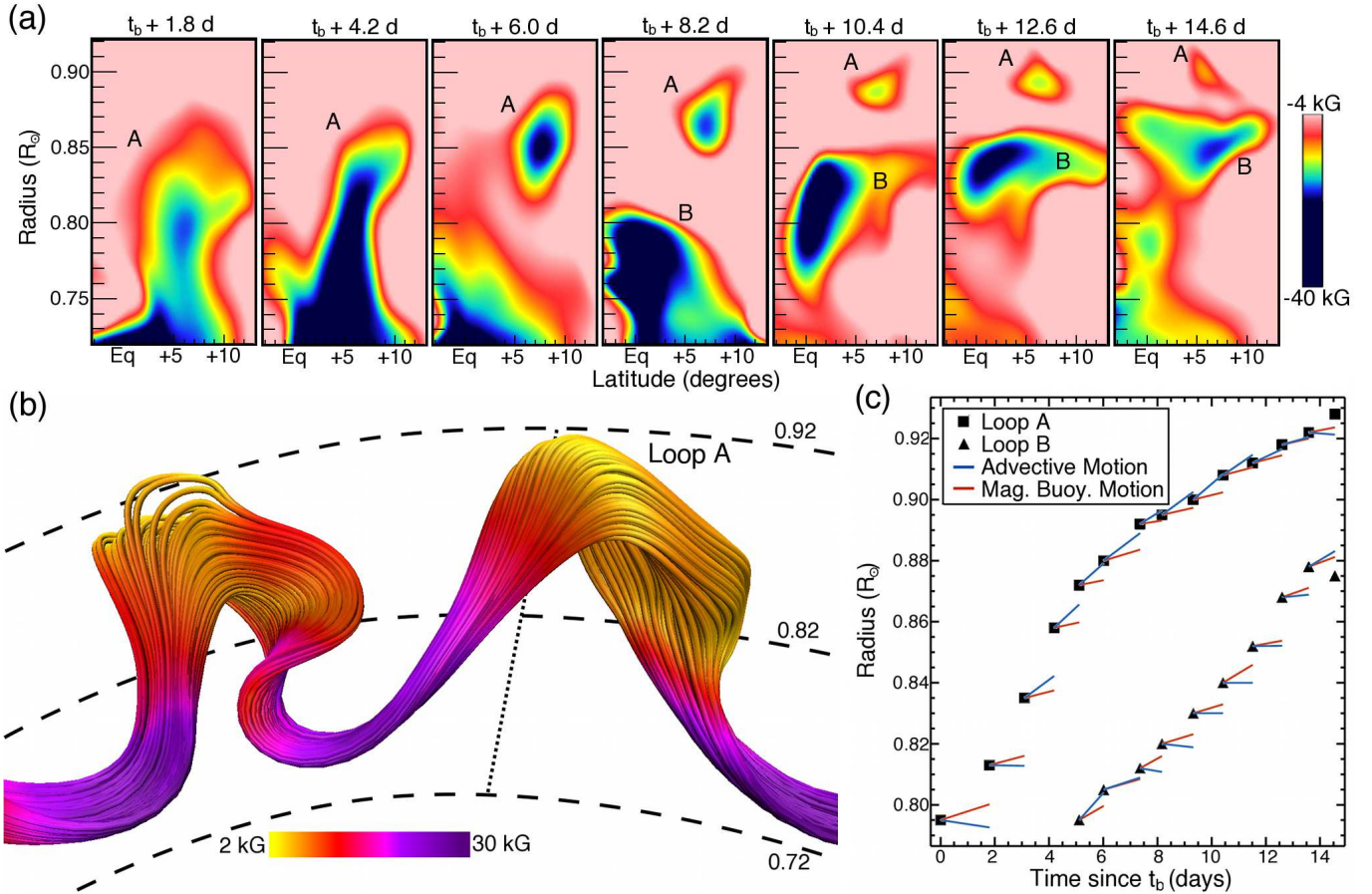}
\caption{ \footnotesize Analyzing a rising loop. 
(a)~2-D cuts in longitude at successive times (tracking in longitude at the local rotation rate of the loop) showing toroidal magnetic field over radius and latitude. 
The rising magnetic loop A is seen in cross-section starting at 0.81$R_\odot$ at $t = t_b$ and rising to 0.91$R_\odot$ after roughly 15 days. 
Proto-loop B is also seen rising starting at 8.6 days, but the top of loop B never rises above 0.88$R_\odot$. 
(b)~3-D visualization of magnetic field lines in the core of a wreath which produces four loops (two shown here, one of which is loop A) at $t_b + 14.6$ days. 
Perspective is looking down along the rotation axis toward the equatorial plane. 
Coloring indicates field magnitude. 
Dashed lines indicate radial position. 
Dotted line shows the cutting plane used in the left-most panel above. 
(c)~Radial location of the top of a buoyant loop as a function of time since $t_b$, along with movement attributable to magnetic buoyancy (red lines) or to advection by convective upflows (blue lines).
\label{rise}}
\end{center}
\end{figure*}

Case S3 exhibits turbulent convective patterns shown in Figure~\ref{meetDS}(a) which are largely vortical at high latitudes and aligned with the rotation axis near the equator. 
The convection builds and maintains a strong differential rotation that is prograde at the equator and retrograde at mid to high latitudes (Figure~\ref{meetDS}(c)). 
This organized shear drives the creation of toroidal magnetic structures at low latitudes in each hemisphere, as demonstrated in B10. 
Here the increased level of turbulence enhances the power in smaller-scale components of the toroidal field $B_\phi$ (Figure~\ref{meetDS}(b)) while still retaining a substantial zonally-averaged toroidal field $\langle B_\phi \rangle$ (Figure~\ref{meetDS}(d)).

In addition to creating strong magnetic structures near the base of the convective region, case S3 also undergoes cycles of magnetic activity and reversals of global magnetic polarity similar to those described in case D5 in B11. 
This is consistent with results from parameter surveys with ASH simulations, which indicate that decreasing both $\nu$ and $\Omega$ can yield cyclic behavior seen at 5$\Omega_\odot$ at lower rotation rates \cite{Bro11a}. 
Because of the large computational cost of the DSMAG SGS model, case S3 was started using a less diffusive descendant of case D3 in B10, which we label case D3b, as initial conditions \cite[see][]{Nel10}. 
Figure \ref{history}(a) shows the temporal evolution of the hemispherical volume-average of toroidal magnetic field, $\left[ B_\phi \right]$, in the progenitor case over approximately 5000 days, demonstrating the irregular cycles this model yields. 
Case S3 continues this behavior over about 1300 simulated days starting from time $t_s$. 
The temporal evolution of $\left[ B_\phi \right]$ in case S3 is shown in Figure~\ref{history}(b), revealing two reversals of global magnetic polarity. 

Some caution should be used in interpreting any LES dynamo simulation, given the potential sensitivity of dynamo action to magnetic dissipation and the nonlinear, nonlocal nature of turbulent magnetic induction, which makes reliable SGS modeling difficult. 
However, we believe the essential large-scale dynamics exhibited in this simulation are robust and are largely insensitive to the SGS model.  
Indeed convective dynamo simulations with differing prescriptions for SGS diffusion exhibit similar large-scale magnetic structures \cite[B10; B11;][]{Ghi10, Rac11}.

Here we will discuss buoyant magnetic structures which coherently rise above 0.90$R_\odot$ while remaining connected to the large-scale toroidal wreaths. 
Using these criteria, we have identified nine buoyant magnetic loops, indicated by hash marks in Figure~\ref{history}(b). 
Eight loops are seen in the northern hemisphere and one in the southern hemisphere. 
We expect that the apparent asymmetry is simply the result of having studied only two magnetic cycles.

\section{Buoyant Magnetic Loops}

Buoyant magnetic loops arise from the cores of toroidal magnetic wreaths near the base of the simulated domain. 
These wreaths have significant $\langle B_\phi \rangle$ components that peak around 5$\mathrm{\, kG}$ while also having strong non-axisymmetric fields. 
Figure~\ref{meetDS}(b) shows a typical $B_\phi$ configuration involving a negative polarity wreath in the northern hemisphere spanning $95^\circ$ in longitude and a positive polarity wreath in the southern hemisphere extending over  $270^\circ$ in longitude. 
As demonstrated in cases D3 (B10) and D5 (B11), these magnetic wreaths are highly nonuniform and display significant internal variation as well as a high degree of connectivity with the rest of the domain. 
In case S3 portions of the wreaths can have coherent cores in which $B_\phi$ can regularly exceed 25$\mathrm{\, kG}$ and have peak values as high as 54$\mathrm{\, kG}$. 
In these cores, bundles of magnetic field lines show very little local connectivity with the rest of the domain or even the other portions of the wreath. 
A single wreath of a given polarity may not form a coherent core at all or may have more than one core, and a single core may produce multiple buoyant loops. 
Of the nine buoyant loops investigated here to rise past 0.90$R_\odot$, one coherent core produces four buoyant loops, another produces three, and two more cores each yield a single buoyant loop.

Some of the coherent wreath cores can become buoyant magnetic loop progenitors or proto-loops. 
In these proto-loops the strong Lorentz forces result in highly suppressed convective motions. 
If we examine extended regions in the cores of wreaths with a local ratio of magnetic to kinetic energy above a fiducial value of 100, we identify at least 35 proto-loops at the times where the nine buoyant loops arise. 
Thus the large majority of proto-loops do not evolve into mature buoyant loops, generally due to unfavorable interactions with convective flows. 
When magnetic field strengths exceed 35$\mathrm{\, kG}$ the proto-loops become significantly underdense as magnetic pressure displaces fluid, causing buoyant acceleration. 
With some rise a proto-loop can enter a region of less suppressed giant cell convection. 
These flows will advect portions of the proto-loop downward at cell edges and upward in the core of the giant cells. 
The rise of the top of a magnetic loop is shown in cross-section by sampling $B_\phi$ roughly every 2 days in Figure~\ref{rise}(a). 
Not all proto-loops become buoyant loops by our criteria. 
For example, loop B in Figures~\ref{rise}(a,c) begins to rise but is prevented from passing 0.88 $R_\odot$ when the top of the loop encounters a strong downflow.

One way to track these buoyant loops is to use 3-D tracings of magnetic field lines using the VAPOR software package \cite{Cly07}. 
In our simulations with finite resistivity, individual field lines do not maintain their identity in time. 
However, one can achieve some measure of consistency as the structure moves and evolves by tracking field line ensembles. 
We track the very strong fields at the bottom of the loops near the base of the domain and we randomly seed large numbers of field lines (here 1000) in those highly magnetized footpoints at each time step. 
Figure~\ref{rise}(b) shows a 3-D rendering of magnetic fields lines for two sample loops near the peak of their rise. 
Similar field line tracings have been studied at various times during the rise of these loops.

At maximum rise, the sample magnetic loop A extends from 0.73$R_\odot$ to 0.93$R_\odot$. 
The magnetic fields exceed 40$\mathrm{\, kG}$ at the base of the loop but become much weaker near the top of the loop, with field strengths as low as 2$\mathrm{\, kG}$. 
Such loops are embedded in the much larger wreaths which have an average cross-sectional area of $13800 \mathrm{\, Mm}^2$. 
The cross-sectional area of loop A is $120 \mathrm{\, Mm}^2$ at 0.795$R_\odot$ and $520 \mathrm{\, Mm}^2$ at its peak radial position of 0.923$R_\odot$. 
Accounting for the continued expansion that would likely occur if this loop were able to rise further, the cross-sectional area is reasonable compared with the typical area of a large sunspot at the solar surface, which is roughly $2500 \mathrm{\, Mm}^2$ \cite{Zwa87}. 
If the loop were rising adiabatically over the same interval, the cross-sectional area should change in inverse proportion to the change in background pressure, which decreases here by a factor of 17.1, rather than the observed expansion by a factor of 4.3. 
The top of the loop must then have a net outflow of heat or material in order to avoid expanding adiabatically. 
The loops show a measurable deficit in density and thermodynamic pressure relative to their surroundings, but they do not possess any detectable signature in temperature or entropy. 
This indicates that they are thermally ``leaky'' and able to equilibrate quickly compared to the timescale for radial motion. 
A simple estimate of the thermal diffusion time across one of these structures at mid-convection zone is on the order of 50 days, implying that there is likely also a divergent flow at the top of the loop, moving fluid along field lines.
We see some evidence for such flows with roughly 1$\mathrm{ \, m \, s}^{-1}$ speeds.

Once a loop has begun to rise, its radial motion is dominated by advection and magnetic buoyancy. 
Figure~\ref{rise}(c) illustrates the motion of loop A which begins to rise buoyantly at $t_b$, while also indicating the components of the motion due to advection and magnetic buoyancy. 
To compare motion due to magnetic buoyancy, we define a magnetic buoyancy velocity $v_{\mathrm{mb}}$ at  the times sampled in Figure~\ref{rise}(c). 
Magnetic buoyancy acceleration is here the fractional density deficit in the loop compared to the average density of the surrounding fluid times the local gravitational acceleration. 
For a magnetic structure in local thermal equilibrium, this reduces to the ratio of magnetic pressure inside the loop to thermodynamic pressure in the surrounding fluid times gravitational acceleration. 
To compute $v_{\mathrm{mb}}$ we integrate the magnetic buoyancy acceleration over the intervals between times plotted in Figure~\ref{rise}(c) (roughly 2 days), which likely provides a lower bound on this velocity. 
The advective velocity $v_{\mathrm{ad}}$ is the volume-averaged velocity of the surrounding fluid. 
The pressure and velocity of the surrounding  fluid are calculated by taking averages over the convective updraft while excluding regions with field magnitude greater than 4$\mathrm{\, kG}$. 
Initially the sample proto-loop experiences an upward $v_{\mathrm{mb}} = 46.1 \mathrm{ \, m \, s}^{-1}$. 
After 3 days of movement dominated by magnetic buoyancy, the loop gets caught in a convective updraft and $v_{\mathrm{ad}}$ becomes greater than $v_{\mathrm{mb}}$. 
Even though advective motions dominate, magnetic buoyancy continues to drive an average upward motion at 32.3$\mathrm{ \, m \, s}^{-1}$ relative to the surrounding fluid. 
Continued buoyant acceleration of the loop as the magnetic pressure weakens is achieved because its density perturbation decreases at roughly the same rate as does the background density stratification. 
Once the top of the loop has entered the main convective upflow it experiences advection at an average velocity of 53.1$\mathrm{ \, m \, s}^{-1}$. 
The presence of magnetic buoyancy forces allow this loop to rise in 14.6 days while the average upflow traverses the same distance in 21.7 days and magnetic buoyancy alone would require 30.6 days.

Additional accelerations are present but not shown, including thermal buoyancy, which is significant early in the rise of the loop, and magnetic tension, which is of the same order of magnitude as the advective motion near maximum radial extent at 14.6 days and helps tether the loop to that height. 
Thermal buoyancy is distinguished from magnetic buoyancy by averaging over the convective updraft but excluding regions with magnetic fields above 4$\mathrm{\, kG}$. 
An additional apparent motion at early times is produced as toroidal magnetic field used to track the loop is converted to radial magnetic field in the sides of the loop.
Because advection plays a crucial role in the transport of these magnetic loops, their size scale is set by the size of the convective giant cells. 
The nine loops studied here have an average extent of $15.4^\circ$ in longitude when measuring across the bottom of the loop, whereas the average distance between convective downflows in the equatorial region is $16.4^\circ$ in longitude. 

\section{Reflections}

In this paper we have presented a 3-D MHD simulation that combines turbulent convection, rotation, and stratification to produce solar-like differential rotation and wreaths of large-scale toroidal magnetic field at the base of the convection zone.
These undergo cycles of magnetic activity and reversals of global magnetic polarity. 
Most notably the wreaths also exhibit buoyant magnetic loops capable of coherently traversing much of the convective layer. 
Such loops can only be realized when the field amplitude in a portion of a wreath exceeds 35$\mathrm{\, kG}$, the diffusion timescale across the proto-loop (here 50 days) is much longer than the timescale for rise due to magnetic buoyancy, and the interactions between rising loops and convective flows are favorable.
These buoyant loops which appear at cycle maximum can have toroidal field strengths of 45$\mathrm{\, kG}$ at their base and 5$\mathrm{\, kG}$ at their top. 
Their size scales are set by the size of the convective giant cells and they have cross-sectional areas at 0.90$R_\odot$ that are reasonable compared to the area of a large sunspot. 

We must be cautious in suggesting that these rising magnetic loops can make it through to the surface of the star.  
Our global simulations here only extend to 0.97$R_\odot$ and currently place an impenetrable boundary there, for we cannot cope with the intense small scales of convection seen as supergranulation and granulation near the surface. 
The presence of the domain boundary deflects all flows, leading to some uncertainty about the fate of the rising loops that could only be resolved by linking flows and magnetism in the upper reaches of ASH to another high-resolution compressible domain closer to the surface. 
This is a task we are now pursuing in parallel with global modeling. 

It is noteworthy that within this simulation convection generates differential rotation which in turn generates toroidal flux which then buoyantly destabilizes and rises.
Each link in this chain is physically well established.
Our primary accomplishment here is to capture all these processes self-consistently within a single simulation.
This represents an essential step toward unifying numerical models of global-scale convective dynamos and surface flux emergence.\\

We thank Kyle Auguston, Christopher Chronopolous and Yuhong Fan for comments and advice. 
This work is supported in part by the NASA grants NNX08AI576G and NNH09AK14I, ERC grant STARS2 207430, and NSF Astronomy and Astrophysics Postdoctoral Fellowship AST 09-02004. 
NCAR is supported by the NSF. 
CSMO is supported by NSF grant PHY 08-21899.

\end{document}